# Diffusion Pore Imaging by Hyperpolarized $^{129}$Xenon Nuclear Magnetic Resonance

Tristan Anselm Kuder[1], Peter Bachert[1], Johannes Windschuh[1], Frederik Bernd Laun[1,2]

[1] *Medical Physics in Radiology, German Cancer Research Center (DKFZ), Heidelberg, Germany*

[2] *Quantitative Imaging Based Disease Characterization, German Cancer Research Center (DKFZ), Heidelberg, Germany*

**Nuclear magnetic resonance (NMR) diffusion measurements are widely used to derive parameters indirectly related to the microstructure of biological tissues and porous media[1-3]. However, a direct imaging of cell or pore shapes and sizes would be of high interest. For a long time, determining pore shapes by NMR diffusion acquisitions seemed impossible, because the necessary phase information could not be preserved[4,5]. Here we demonstrate experimentally using the measurement technique which we have recently proposed theoretically[6] that the shape of arbitrary closed pores can be imaged by diffusion acquisitions, which yield the phase information. For this purpose, we use hyperpolarized xenon gas in well-defined geometries. The signal can be collected from the whole sample which mainly eliminates the problem of vanishing signal at increasing resolution of conventional NMR imaging. This could be used to non-invasively gain structural information inaccessible so far such as pore or cell shapes, cell density or axon integrity.**

To investigate the structural properties of biological tissues and porous media, mainly the sizes and shapes of the cells or pores are of interest and not their positions. Regarding the achievable resolution, besides the required gradient strength, the main limiting factor for conventional NMR imaging is the remaining signal-to-noise ratio (SNR). NMR diffusion measurements can overcome this limitation by using the signal from a larger sample volume yielding information averaged over all contained cells or pores[3,4,7-9].

In NMR diffusion measurements[10], to detect the Brownian motion of diffusing spins, the static magnetic field $B_0$ is modified by a field gradient profile $\mathbf{G}(t)$, which depends on the time $t$ and fulfils $\int_0^T dt\, \mathbf{G}(t) = 0$ with the total gradient duration $T$ resulting in the field strength $B(t) = B_0 + \mathbf{G}(t) \cdot \mathbf{x}(t)$ for a particle at $\mathbf{x}(t)$. The signal attenuation due to the diffusion weighting gradients is given by $S = \langle \exp(i\phi) \rangle$, where $\langle \cdot \rangle$ denotes the average over all possible random paths, and $\phi = -\gamma \int_0^T dt\, \mathbf{G}(t) \cdot \mathbf{x}(t)$ is the phase accumulated during the random walk $\mathbf{x}(t)$ with the gyromagnetic ratio $\gamma$.

To gain information about diffusion restrictions, a particularly useful technique is q-space imaging[4,11,12]. Here, two short gradient pulses of duration $\delta$ are applied at $t = 0$ and $t = T - \delta$ with the gradient vectors $\mathbf{G}$ and $-\mathbf{G}$. It is assumed that the diffusive motion during $\delta$ can be neglected. Defining the q-value $\mathbf{q} = \gamma \delta \mathbf{G}$ and the pore space function $\rho(\mathbf{x})$, which equals the reciprocal pore volume inside the pore and zero outside, and assuming that $T$ is sufficiently long so that the correlations between the starting position $\mathbf{x}_1 = \mathbf{x}(0)$ and the final position $\mathbf{x}_2 = \mathbf{x}(T)$ vanish, the signal attenuation becomes

$$S_2(\mathbf{q}) = \left\langle e^{i\mathbf{q}\cdot(\mathbf{x}_2 - \mathbf{x}_1)} \right\rangle = \int_V d\mathbf{x}_2\, \rho(\mathbf{x}_2) e^{i\mathbf{q}\cdot\mathbf{x}_2} \int_V d\mathbf{x}_1\, \rho(\mathbf{x}_1) e^{-i\mathbf{q}\cdot\mathbf{x}_1} = |\tilde{\rho}(\mathbf{q})|^2. \qquad [1]$$

$\tilde{\rho}(\mathbf{q})$ symbolizes the Fourier transform (FT) of $\rho(\mathbf{x})$ and $V$ is the volume where the pore is located. Thus, in analogy to diffraction experiments, the magnitude can be observed but the inverse FT cannot be performed due to the missing phase information.

Our approach[6], which we demonstrate experimentally here, uses a combination of a long and a short gradient pulse (Fig. 1a, gradient shape 1) resulting in the gradient profile $\mathbf{G}(t) = \mathbf{G}_1 = -\delta_2/\delta_1 \mathbf{G}_2$ for $0 \leq t \leq \delta_1$ and $\mathbf{G}(t) = \mathbf{G}_2$ for $\delta_1 < t \leq T$ with the q-value $\mathbf{q} = \gamma \delta_2 \mathbf{G}_2$ and $\delta_1 + \delta_2 = T$. $\mathbf{G}_1$ and $\mathbf{G}_2$ are the gradient vectors. Consequently, the signal is given by[6,13,14]

$$S(\mathbf{q}) = \left\langle \exp\left[i\mathbf{q}\cdot\left(\frac{1}{\delta_1}\int_0^{\delta_1} dt\, \mathbf{x}(t) - \frac{1}{\delta_2}\int_{\delta_1}^T dt\, \mathbf{x}(t)\right)\right] \right\rangle = \left\langle e^{i\mathbf{q}\cdot(\mathbf{x}_{CM,1} - \mathbf{x}_{CM,2})} \right\rangle. \qquad [2]$$

$\mathbf{x}_{CM,1}$ and $\mathbf{x}_{CM,2}$ denote the centers of mass of the particle random walks during the gradients $\mathbf{G}_1$ and $\mathbf{G}_2$. Each of these gradients imprints a phase that a particle resting at $\mathbf{x}_{CM,1}$ or $\mathbf{x}_{CM,2}$ would acquire. In the long-time limit (long $\delta_1$) $\mathbf{x}_{CM,1}$ of each random walk during the first gradient converges to the pore center of mass $\mathbf{x}_{CM}$. Neglecting the diffusive displacements during the short period $\delta_2$, $\mathbf{x}_{CM,2}$ converges to the final point $\mathbf{x}_2$ of the random walk and Eq. 2 becomes

$$S(\mathbf{q}) = \left\langle e^{i\mathbf{q}\cdot(\mathbf{x}_{CM}-\mathbf{x}_2)} \right\rangle = e^{i\mathbf{q}\cdot\mathbf{x}_{CM}} \int_V d\mathbf{x}_2\, \rho(\mathbf{x}_2) e^{-i\mathbf{q}\cdot\mathbf{x}_2} = e^{i\mathbf{q}\cdot\mathbf{x}_{CM}} \tilde{\rho}(\mathbf{q}). \qquad [3]$$

Thus, the phase information is preserved and the inverse FT can be performed to obtain the pore shape. The long gradient results in the factor $e^{i\mathbf{q}\cdot\mathbf{x}_{CM}}$, which shifts the center of mass of all pores to a common point, while the short gradient acts as an imaging gradient. Consequently, in a medium with many similar pores, all pores contribute to one reconstructed image and an average pore space function can be measured at a higher SNR compared to conventional NMR imaging.

The actual temporal shape of the gradient pulses is not of essential importance, as long as one short and at least one long gradient pulse are present[13]. This allows using a spin echo sequence to compensate for the signal loss due to local field inhomogeneities (Fig. 1a, gradient shape 2). To demonstrate the influence of the gradient shape on the measured signal, for the experiments presented here, additionally to gradient shape 2, the modified gradient shapes 3 and 4 were used.

Experiments were performed using gas filled phantoms containing parallel plates and triangular shaped pores (Fig. 1b). Using gas diffusion is advantageous for the experimental demonstration compared to water diffusion since the diffusion coefficient is several orders of magnitude higher (see Methods) and thus the long-time limit can be reached in pores on the millimeter scale, which are easy to realize in phantoms. Despite the high diffusion coefficient, the maximum gradient amplitude of a clinical MR scanner was sufficient. Due to the low NMR signal of thermally polarized gases, hyperpolarized $^{129}$Xe gas was used, which was generated by spin exchange optical pumping (SEOP)[15-17]. For technical reasons[15], a mixture of helium, nitrogen and xenon gas was used, which was transferred continuously flowing to a pumping cell, where it was mixed with rubidium vapor (Fig. 1c). By laser optical pumping, an electron-spin polarization of the rubidium atoms was achieved, which was transferred to the xenon nuclear spins by spin exchange collisions. The hyperpolarized gas was then used for the phantom measurements in the MR scanner.

Figure 2 shows the signal acquired using one gradient direction (gradient shape 3 in Fig. 1a) orthogonal to parallel plates as well as the inverse FT for different durations $\delta_1$ of the long gradients. The measured signal and its FT (dots) are in good agreement with the simulation of the diffusion process (sold lines), where the Bloch-Torrey equation was solved numerically using a matrix approach[5,18]. Due to the symmetry of the slits, the imaginary part of the signal is zero[13,19,20]. However, contrary to q-space imaging, the negative signal values allow the reconstruction of the pore space function. For $\delta_1 = 440\,\text{ms}$, the FT of the oscillating signal clearly shows the single-slit function, which comprises the signal of all slits in the phantom. Two deviations from the ideal rectangular single-slit function due to the finite gradient durations can be observed, which were not considered in Eq. 3. The finite $\delta_1$ causes a blurred $\mathbf{x}_{CM,1}$ (Eq. 2) and a finite slope at the edges. Since $\delta_2$ is not infinitesimal, the edge enhancement effect[13,21] can be observed (Fig. 2a, blue arrow): $\mathbf{x}_{CM,2}$ deviates from the trajectory endpoint $\mathbf{x}_2$ and thus the signal is effectively pushed away from the domain edges yielding an increased signal near the boundaries. For shorter $\delta_1$ (Fig. 2b,c), the distribution of $\mathbf{x}_{CM,1}$ gradually broadens. Consequently, for $\delta_1 = 80\,\text{ms}$, the signal oscillations and the pore shape information are mostly lost, which shows the importance of reaching the long-time limit.

Figure 3 demonstrates the possibility to image an average pore space function using two different plate distances in a single phantom. The contributions of different pore sizes can be clearly observed. The finite gradient durations cause deviations from a superposition of a narrow and a wide rectangular function.

The measurements of equilateral triangular pores in Fig. 4 exhibit good agreement with the simulations. As this domain is not point-symmetric[13,20], non-zero imaginary parts can be observed for the vertical gradient direction (arrows). Thus, the full phase information is obtained, which is needed to reconstruct arbitrary pores shapes. The saw-tooth shaped FT in Fig. 4a corresponds to the expected projection of the pore on the gradient direction. The need to use a long-narrow gradient scheme is demonstrated in Fig. 4b,c. If the long-narrow requirement is violated, the information about the pore shape is gradually lost. For the horizontal gradient direction (Fig. 4d), the signal is real due to the symmetry plane orthogonal to the gradient vector.

In Fig. 5a, pore imaging measurements for the phantom with 170 triangular domains are shown. The data were acquired radially with 19 gradient directions and 15 q-values using gradient shape 2. The triangle is clearly visible and in very good agreement with the simulations. All pores in the phantom contribute to one average pore image. For shorter $\delta_1$, the image is blurred due to the broadening of the $\mathbf{x}_{CM,1}$ distribution; because of the edge enhancement effect, the triangle appears shrunk for prolonged $\delta_2$. For comparison, Fig. 5b shows pore images measured and simulated for gradient shape 3. A detailed comparison of the influence of the gradient shapes can be seen in Fig. 5c, resulting in small differences of the measured signals, which can be reproduced remarkably well by the simulation of the diffusion process. Consequently, the gradient shape can be optimized for applications. For example, gradient shape 3 could be used to reduce imaging artifacts caused by concomitant magnetic fields[22].

In conclusion, we have demonstrated experimentally that the proposed new form of diffusion based magnetic resonance imaging can reveal the average shape of arbitrary closed pores in a volume with many pores, which mainly overcomes the SNR limitation of conventional NMR imaging[23]. Regarding practical applications, namely to image cells, several complications would have to be addressed as the high required gradient amplitudes, phase stability, relaxa-tion time requirements, permeability and size distributions. Several limitations were discussed in detail recently[13].

Diffusion pore imaging is currently a quickly evolving field of research, which is now in the stage of first experimental verifications. Very recently, for cylindrical capillaries and the long-narrow gradient scheme, initial pore images with micron resolution were presented[24]. A different approach to diffusion pore imaging[25], which however requires the pore to be point-symmetric[20], was demonstrated by application to capillaries. For point-symmetric domains, pore imaging techniques rely on the sign of $\tilde{\rho}(\mathbf{q})$, but the imaginary part vanishes. Our results show the possibility to acquire the phase information of $\tilde{\rho}(\mathbf{q})$ and verify experimentally that arbitrary non point-symmetric pores can be imaged without relying on assumptions, which is desirable for practical applications on the microscopic scale. Besides imaging of porous media such as oil containing rocks or cement, gaining information about cells in biological tissues, thus e.g. retrieving parameters usually derived from histology as tumor cellularity or axon integrity could be an eventual aim.

**Methods**

**Experimental setup**

To generate the hyperpolarized gas, a mixture containing 0.95 % xenon gas and the buffer gases helium and nitrogen was used. The xenon gas contained the NMR visible isotope $^{129}$Xe ($I = 1/2$) in natural abundance. In the optical pumping cell placed in an external magnetic field $B_H$, rubidium vapor was generated by external heating (Fig. 1c). Circularly polarized laser light was provided by a Coherent FAP Duo Laser (60 W, 795 nm) in combination with a polarizing beam splitter and quarter wave plates and was used to selectively populate an electron-spin state[15]. The gas mixture was polarized continuously flowing through the pumping cell at 2 bar absolute pressure, which was reduced to atmospheric pressure by a nonmagnetic valve after passing through the cell. The gas was transferred at a small constant flow (ap-proximately 100 ml/min) to the phantom in the clinical MR scanner (Siemens Avanto, 1.5 T static magnetic field, 40 mT/m maximum gradient strength). The flow was directed orthogonally to the gradient directions and thus did not influence the diffusion weighted NMR signal, which was additionally verified by varying the flow rate. The time between two 90° pulses was 20 s so that a considerable amount of polarization was restored due to the gas replacement.

The phantoms built of acrylic glass were orientated such that the plates and the axes of the triangular cutouts were parallel to the static magnetic field in the MR scanner. Slice selective excitation (slice thickness 4 cm) and refocusing pulses located in the plate or triangle area were employed. The diffusion weighted signal was normalized to the signal intensity acquired immediately after the excitation pulse to account for polarization fluctuations. The free diffusion coefficient at room temperature needed for the simulations of the diffusion process was estimated to $D_0 = (37000 \pm 2000)\,\mu m^2/ms$ using a conventional diffusion weighted measurement sequence in a phantom without barriers. For comparison, the diffusion coefficient is approximately $6000\,\mu m^2/ms$ for pure xenon gas and $2\,\mu m^2/ms$ for free water.

**Simulations**

For the simulation of the diffusion process, analytically calculated eigenfunctions and eigenvalues of the Laplace operator with the respective boundary conditions[5,13] of the domains were employed to calculate the diffusion weighted signal for the considered gradient shapes. This matrix approach, which was implemented in Matlab (MathWorks), allows a more accurate and efficient approximate solution of the Bloch-Torrey equation than Monte Carlo simulations[5]. The above mentioned value $D_0$ and 50 eigenvalues were used.

**References**

1. Callaghan, P. T. *Principles of Nuclear Magnetic Resonance Microscopy* (Oxford University Press, Oxford England, 1991).

2. Sen, P. N. Time-dependent diffusion coefficient as a probe of geometry. *Concepts Magn. Reson. Part A* **23A**, 1-21 (2004).

3. Le Bihan, D. Looking into the functional architecture of the brain with diffusion MRI. *Nat. Rev. Neurosci.* **4**, 469-480 (2003).

4. Callaghan, P. T., Coy, A., Macgowan, D., Packer, K. J. & Zelaya, F. O. Diffraction-like effects in NMR diffusion studies of fluids in porous solids. *Nature* **351**, 467-469 (1991).

5. Grebenkov, D. S. NMR survey of reflected Brownian motion. *Rev. Mod. Phys.* **79**, 1077-1137 (2007).

6. Laun, F. B., Kuder, T. A., Semmler, W. & Stieltjes, B. Determination of the defining boundary in nuclear magnetic resonance diffusion experiments. *Phys. Rev. Lett.* **107**, 048102 (2011).

7. Le Bihan, D. et al. Diffusion tensor imaging: concepts and applications. *J. Magn. Reson. Imaging* **13**, 534-46 (2001).

8. Basser, P. J. Inferring microstructural features and the physiological state of tissues from diffusion-weighted images. *NMR Biomed.* **8**, 333-44 (1995).

9. Novikov, D. S., Fieremans, E., Jensen, J. H. & Helpern, J. A. Random walks with barriers. *Nat. Phys.* **7**, 508-514 (2011).

10. Stejskal, E. O. Use of spin echoes in a pulsed magnetic-field gradient to study anisotropic, restricted diffusion and flow. *J. Chem. Phys.* **43**, 3597-3603 (1965).

11. Callaghan, P. T., Macgowan, D., Packer, K. J. & Zelaya, F. O. High-resolution q-space imaging in porous structures. *J. Magn. Reson.* **90**, 177-182 (1990).

12. Cory, D. G. & Garroway, A. N. Measurement of translational displacement probabilities by NMR: An indicator of compartmentation. *Magn. Reson. Med.* **14**, 435-444 (1990).

13. Laun, F. B., Kuder, T. A., Wetscherek, A., Stieltjes, B. & Semmler, W. NMR-based diffusion pore imaging. *Phys. Rev. E* **86**, 021906 (2012).

14. Mitra, P. P. & Halperin, B. I. Effects of finite gradient-pulse widths in pulsed-field-gradient diffusion measurements. *J. Magn. Reson. A* **113**, 94-101 (1995).

15. Walker, T. G. & Happer, W. Spin-exchange optical pumping of noble-gas nuclei. *Rev. Mod. Phys.* **69**, 629-642 (1997).

16. Oros, A. M. & Shah, N. J. Hyperpolarized xenon in NMR and MRI. *Phys. Med. Biol.* **49**, R105-R153 (2004).

17. Shah, N. J. et al. Measurement of rubidium and xenon absolute polarization at high temperatures as a means of improved production of hyperpolarized Xe-129. *NMR Biomed.* **13**, 214-219 (2000).

18. Axelrod, S. & Sen, P. N. Nuclear magnetic resonance spin echoes for restricted diffusion in an inhomogeneous field: Methods and asymptotic regimes. *J. Chem. Phys.* **114**, 6878-6895 (2001).

19. Özarslan, E. & Basser, P. J. MR diffusion-"diffraction" phenomenon in multi-pulse-field-gradient experiments. *J. Magn. Reson.* **188**, 285-294 (2007).

20. Kuder, T. A. & Laun, F. B. NMR-based diffusion pore imaging by double wave vector measurements. *Magn. Reson. Med.*, DOI: 10.1002/mrm.24515 (2012).

21. De Swiet, T. M. Diffusive edge enhancement in imaging. *J. Magn. Reson. B* **109**, 12-18 (1995).

22. Bernstein, M. A. et al. Concomitant gradient terms in phase contrast MR: Analysis and correction. *Magn. Reson. Med.* **39**, 300-308 (1998).

23. Kuder, T. A. & Laun, F. B. Diffusion pore imaging by double wave vector measurements. *11th International Bologna Conference on Magnetic Resonance in Porous Media (Guildford, UK)* **Talk O21** (2012).

24. Hertel, S. A., Hunter, M. & Galvosas, P. Long-narrow PGSE NMR: Evolution of a pulse sequence. *11th International Bologna Conference on Magnetic Resonance in Porous Media (Guildford, UK)* **Poster 6.8** (2012).

25. Shemesh, N., Westin, C. F. & Cohen, Y. Magnetic resonance imaging by synergistic diffusion-diffraction patterns. *Phys. Rev. Lett.* **108**, 058103 (2012).

**Figure 1**

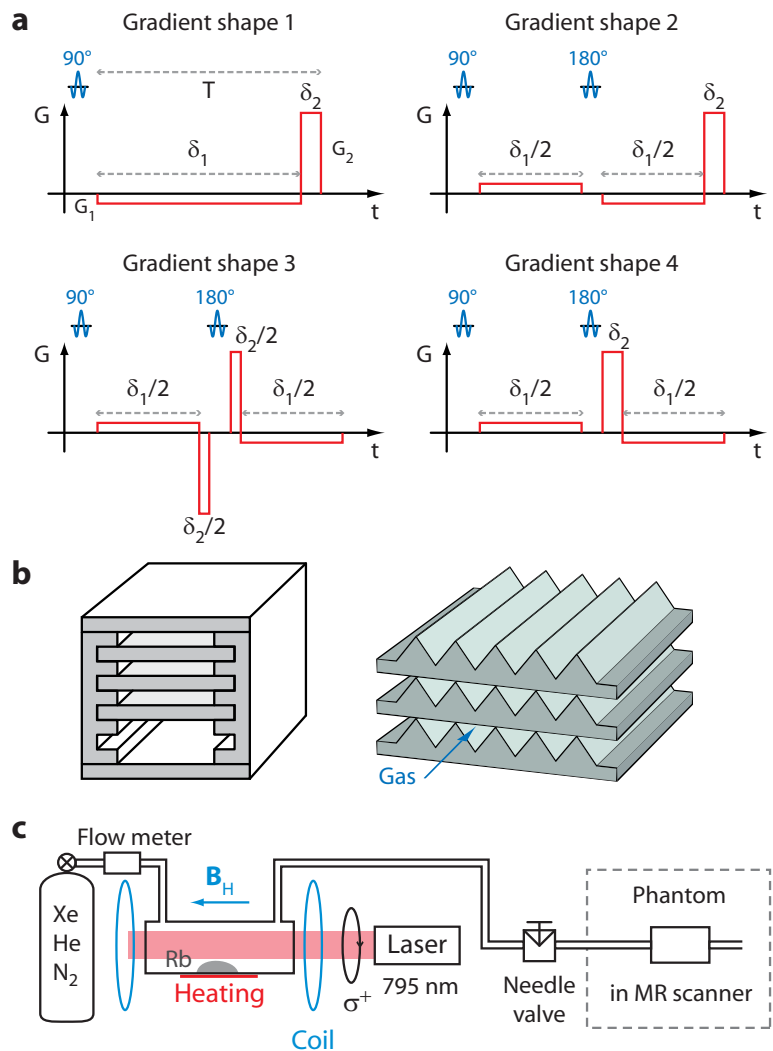

**Schematic representation of the diffusion weighting gradient profiles and of the experimental setup. a**, The long-narrow gradient scheme (shape 1) uses a long diffusion weighting gradient to generate a phase proportional to the pore center of mass and a short imaging gradient. Additionally to the 90° excitation pulse, a 180° refocusing pulse (duration 2.6 ms) can be used to realize a spin-echo sequence (shape 2). The gradient timing can be modified (shapes 3,4). Shapes 2-4 were used for the experiments. **b**, Phantoms: parallel plates (distance: 1 mm, 3 mm, 5 mm) and pores with triangular cross-section (equilateral, edge length 3.4 mm). **c**, Generation of hyperpolarized xenon gas by spin exchange optical pumping.

**Figure 2**

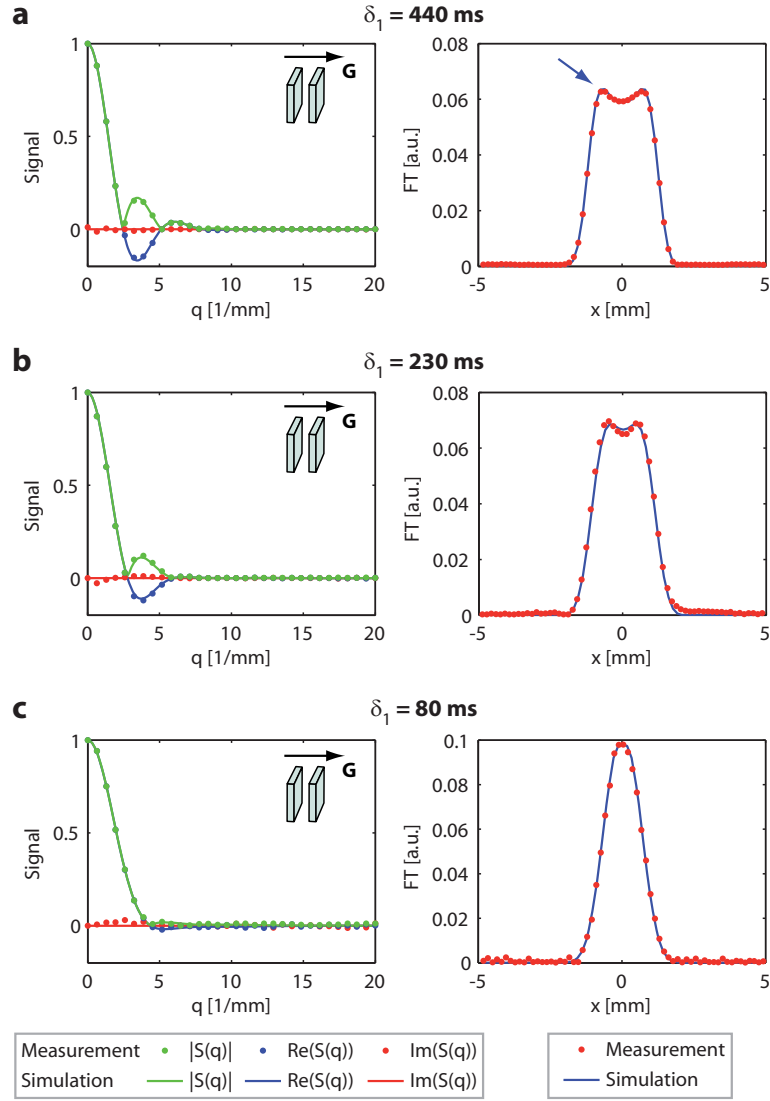

**Measurement results for parallel plates for different durations $\delta_1$ of the long gradient.** The measured signals (dots, left) are in good agreement with the simulation (solid line) of the diffusion process for the parallel plate phantom (distance 3 mm, 10 slits, $\delta_2 = 7$ ms, free diffusion coefficient $D_0 = 37\,000\,\mu m^2/ms$, gradient shape 3). The absolute value of the FT of the signal on the right side clearly shows the single-slit function (averaged over all slits) for the longest diffusion time (**a**, $\delta_1 = 440$ ms). The edge enhancement effect is clearly visible (blue arrow). For shorter diffusion times, the information about the pore shape slowly disappears (**b,c**).

**Figure 3**

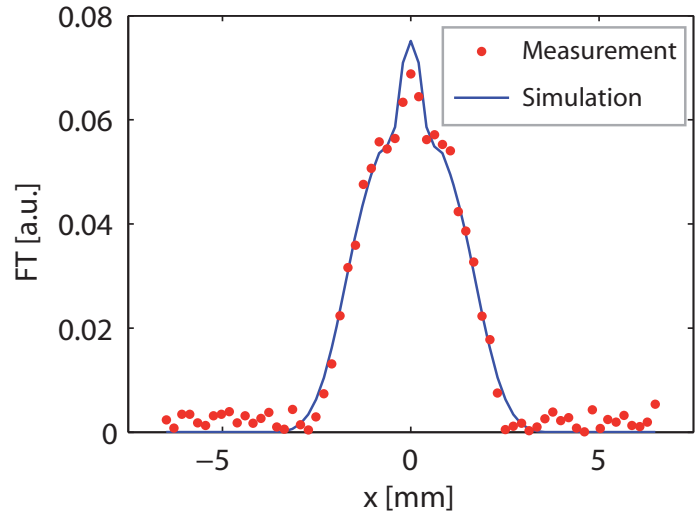

**Combination of different plate distances in one phantom yielding an average pore shape.** The measured and simulated absolute values of the FT for a phantom with different plate distances (six 5-mm slits and two 1-mm slits) both show a superposition of the two pore shapes ($\delta_1 = 440$ ms, $\delta_2 = 5.2$ ms, gradient shape 3, maximum q-value $q_{max} = 15$ mm$^{-1}$). Small differences between measurement and simulation regarding the contributions of the two different pore sizes can be attributed to variations of the gas polarization in the two regions.

**Figure 4**

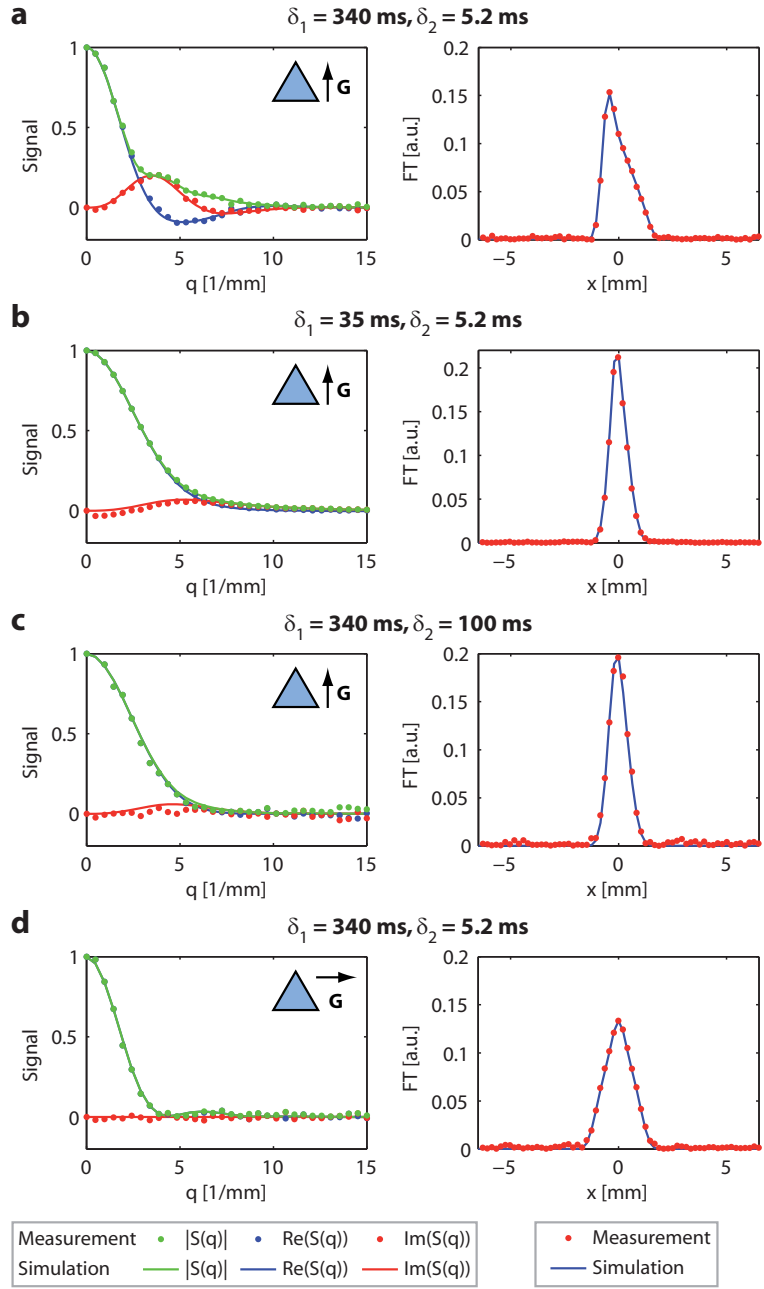

**Measurement results for the equilateral triangular pore shape for single gradient directions. a**, For the vertical gradient direction (arrow), a complex signal arises since there is no mirror plane orthogonal to the gradient direction. The absolute value of the FT of the signal (right side) yields the projection of the pore shape on the gradient direction. **b**, A relatively short first gradient or, **c**, a long second gradient suppresses the oscillations and the pore shape is lost, because the preconditions of Eq. 3 are not met. **d**, For the horizontal gradient direction, the FT again yields the projection of the pore shape (Parameters: edge length $L = 3.4$ mm gradient shape 3, 10 plates each with 17 triangular cutouts).

**Figure 5**

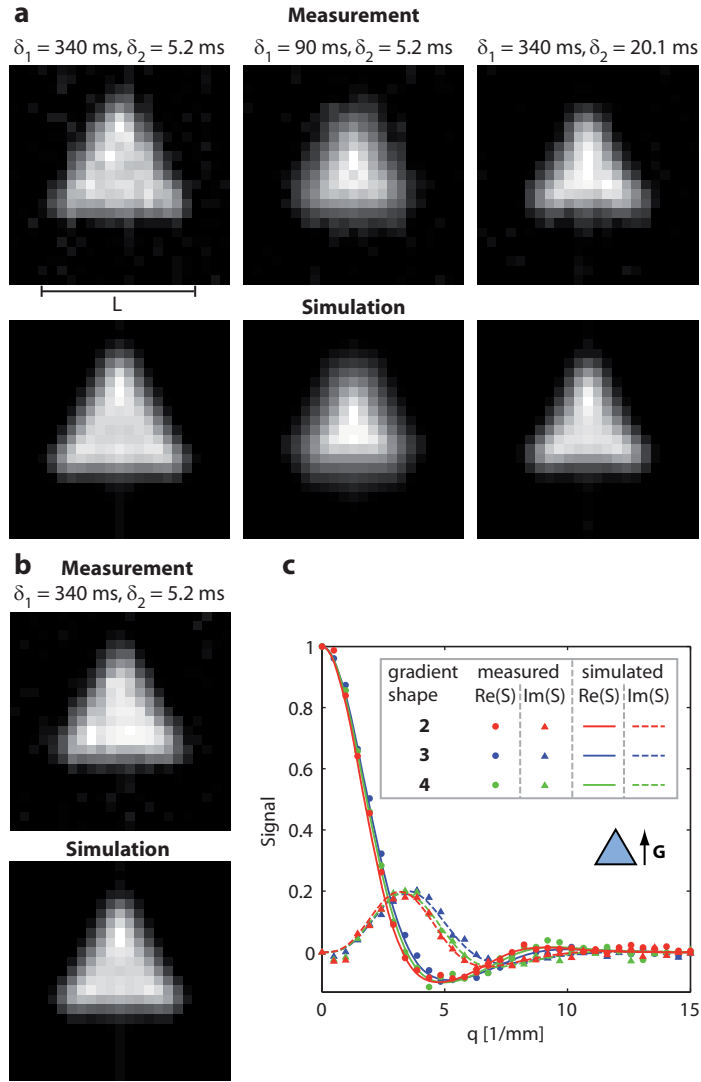

**Pore imaging results for triangular domains. a**, Measurements and simulations for different diffusion times and durations of the short gradient for gradient shape 2 (radial acquisitions with 15 q-values for each of the 19 directions, $q_{max} = 15$ mm$^{-1}$, edge length $L = 3.4$ mm ). All triangular pores in the phantom contribute to one image and thus the SNR limitation of conventional NMR imaging is lifted. **b**, Pore images can also be acquired using gradient shape 3. **c**, The choice of the gradient shape has a small influence on the signal, which can be reproduced by the simulations.